\documentclass[12pt,brazil,english,aps,pre]{revtex4-1}
\usepackage{graphicx}
\usepackage{graphics}
\usepackage{indentfirst}
\usepackage{subfigure}
\usepackage{amsmath, amsthm, amssymb}

\begin{document}

\preprint{This line only printed with preprint option}

\title{Turing patterns and apparent competition in predator-prey food webs
on networks}

\author{L.~D. Fernandes and M.~A.~M.~de Aguiar}

\affiliation{Instituto de F\'{\i}sica `Gleb Wataghin', Universidade
Estadual de Campinas (UNICAMP)\\ 13083-970, Campinas, Brazil}

\begin{abstract}

Reaction-diffusion systems may lead to the formation of steady state
heterogeneous spatial patterns, known as Turing patterns. Their
mathematical formulation is important for the study of pattern formation
in general and play central roles in many fields of biology, such as
ecology and morphogenesis. In the present study we focus on the role of
Turing patterns in describing the abundance distribution of predator and
prey species distributed in patches in a scale free network structure. We
extend the original model proposed by Nakao and Mikhailov by considering
food chains with several interacting pairs of preys and predators. We 
identify patterns of species distribution displaying high degrees of
apparent competition driven by Turing instabilities. Our results provide
further indication that differences in abundance distribution among
patches may be, at least in part, due to self organized Turing patterns,
and not necessarily to intrinsic environmental heterogeneity.

\end{abstract}

\maketitle

\section{Introduction}

Reaction-diffusion systems, in which two or more species interact 
locally and diffuse through the medium, have long been focus of
studies in many different fields, such as Physics, Chemistry and Biology.
Part of the interest in these systems is related to their potential
to form self-organized spatio-temporal patterns, like traveling and
spiral waves \cite{murray_livro_v1} or stationary patterns, called Turing
patterns \cite{murray_livro}.

Working on the problem of morphogenesis \cite{turing} Turing
derived general analytical conditions for the formation of stationary
patterns in reaction-diffusion systems under a mechanism today called 
diffusion driven instability (or Turing instability). Despite the
specific nature of the original problem, the work led to a large number
of applications in chemistry and biology, both theoretical
\cite{murray_artigo1,murray_leopard,murray_artigo2,meinhardt,kishimoto,
baurmann} and, more recently, empirical \cite{bansagi,yamaguchi,sawai}.
A key theoretical contribution was provided by Mimura and Murray
\cite{murray_artigo2}, who applied Turing's idea to understand patchiness
in continuously distributed predator-prey populations.

Recently, Nakao and Mikhailov \cite{nakao} proposed a discrete version of
the prey-predator model of Mimura and Murray \cite{murray_artigo2} in
which the species are organized in patches, instead of being
continuously distributed in space. The patches are represented by nodes of
a complex network such that predators and preys interact locally in each
patch and diffusion occurs through connected nodes. The Turing patterns
obtained in \cite{nakao} present significant differences when compared to
the ones obtained in the analogous system which considers space as a
continuous medium.

In the present work, we extend of the model of Nakao and Mikhailov
\cite{nakao} by considering food chains with more than two species.
We study the dynamics of several pairs of preys and predators that
interact by consuming common preys. We show that the Turing patterns of
population density displayed by the system present nontrivial correlations
in the abundance distributions. In particular, we observe the emergence of
strong competition between preys of adjacent species in the food chain,
despite the fact that no direct competition between them are included in
the equations. These correlations are strictly related to diffusion and
correspond to a new mechanism of apparent competition, driven by Turing
instabilities instead of local interactions. We characterize these
patterns using numerical simulations and mean field approximations. We
also discuss the relevance of these results to patterns of species
distribution in real trophic systems.

\section{Dynamical model}

\begin{figure}[ht]
\centering
\includegraphics[height=5cm]{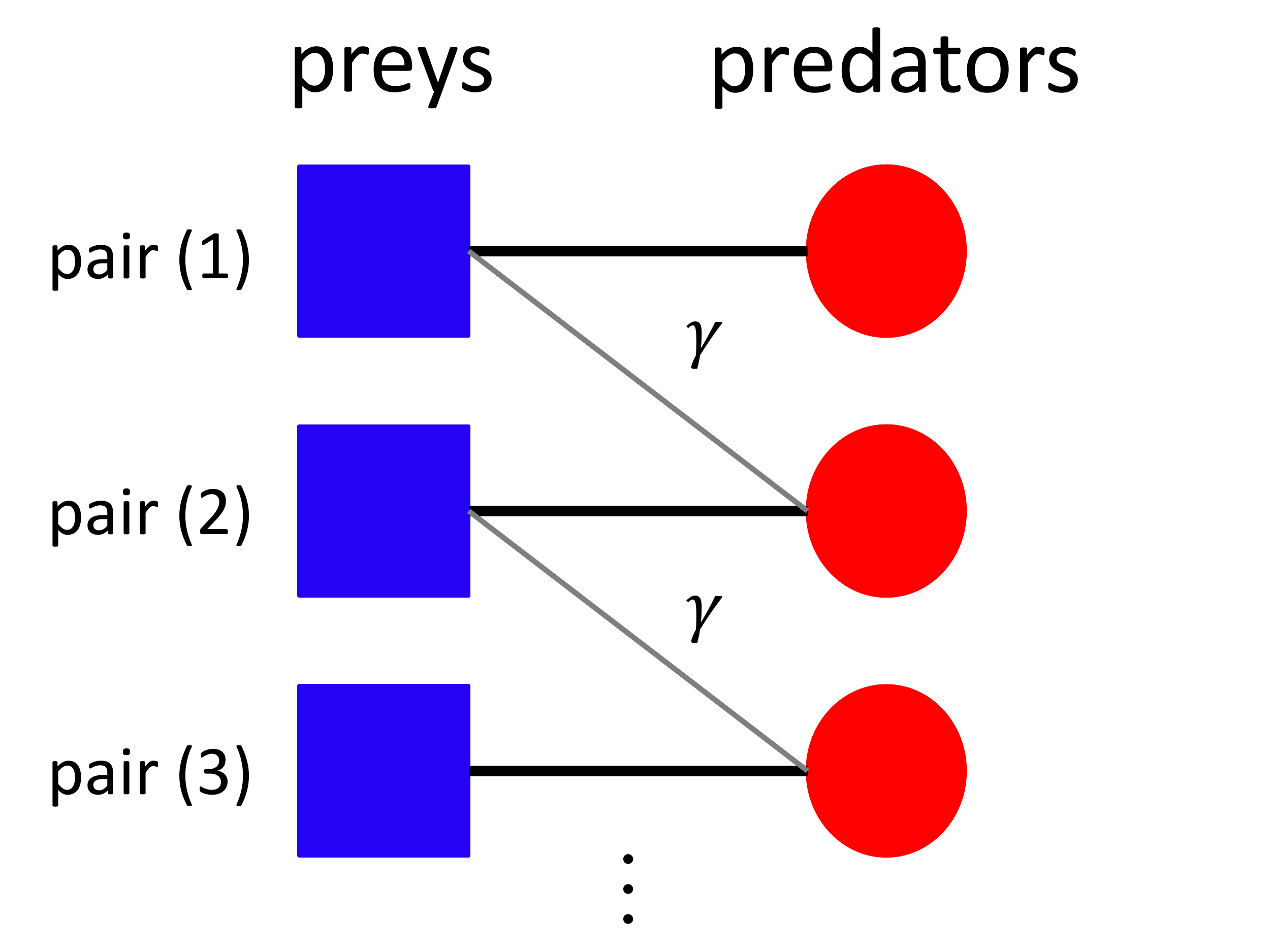}
\caption {(Color online) Hierarchical food chain with three pairs of preys and predators.
Each predator is linked to the previous prey (secondary predation) with 
strength $\gamma$.}
\label{fig1}
\end{figure}

In order to consider more complex reaction diffusion systems we extend
the model introduced by Nakao and Mikhailov \cite{nakao} to food chains
composed by several species of preys and predators. We assume that each
prey species has a primary predator associated to it, forming a pair. The
pairs in the food chain are hierarchically coupled by secondary predation
relations. Thus, the prey in the first pair is consumed by its main
predator and also by the predator in the second pair, though with the
lower intensity $\gamma$. Similarly, the prey of the second pair is
consumed primarily by its associated predator and also by the predator of
the third pair, and so on. Only the last species of prey in this ordered
chain is consumed exclusively by its main predator as, illustrated by the
diagram in figure \ref{fig1}.

The environment where these interactions take place consists of a network
of patches. Species-species interactions, as described by the food chain,
occur locally in each patch and the coupling between patches is
exclusively due to diffusion, which is possible if the patches are
connected in the network. 

The equations describing this dynamical system are given by:

\begin{align}
\frac{d}{dt}u_i^{(l)}(t)&=f(u_i^{(l)},v_i^{(l)})-\gamma
u_i^{(l)}v_i^{(l+1)}+\epsilon\displaystyle\sum_j L_{ij}u_j^{(l)} \notag
\\	
\frac{d}{dt}v_i^{(l)}(t)&=g(u_i^{(l)},v_i^{(l)})+\phi\gamma
u_i^{(l-1)}v_i^{(l)}+\sigma\epsilon\displaystyle\sum_j L_{ij}v_j^{(l)},
\label{eq1}
\end{align}
				
\noindent where $u_i^{(l)}(t)$ and $v_i^{(l)}(t)$ represent the
populations of preys and predators at time $t$ respectively. The label
$l=1,2,3,\cdots$ index the prey-predator pair. 

The functions $f$ and $g$ describe the local interaction between 
preys and predators of each pair (also called {\it reaction functions}).
The terms proportional to $\gamma$ represent the secondary predation
relations between adjacent pairs and the parameter $\phi$ accounts for the
ratio between predator gain and prey loss in the secondary interaction.

The parameters $\epsilon$ and $\sigma$ are, respectively, the prey
mobility and the ratio between predator and prey mobilities. The matrix
$L$ stands for the {\it Laplacian matrix} and accounts for the diffusion
of populations across connected sites. For undirected networks $L$ is
symmetric with $L_{ij}=A_{ij}-k_i\delta_{ij}$, where $A$ is the {\it
Adjacency Matrix} and $k_i$ the degree of node $i$. The adjacency
matrix defines the topology of the network and is given by $A_{ij}=1$
if nodes $i$ and $j$ are connected and $A_{ij}=0$ if they are not. The
degree $k_i = \sum_j A_{ij}$ is the number of connections of node $i$.

The term $\sum_j L_{ij}u_j^{(l)}$ in eq.\ref{eq1} controls the diffusion
of preys $u^{(l)}$. It gives the difference between the total population
of preys $u^{(l)}$ in the sites connected to $i$ and $k_i$ times the
population in the site $i$. If $u^{(l)}$ is the same in all sites the sum
adds to zero and there is no diffusion. A similar term controls the
diffusion of predators in the equation for $v^{(l)}$.

As a simplification, we consider that the intrinsic growth rate of all 
prey species are the same, as is the intrinsic death rate of all predator 
species. In that manner, the functions $f$ and $g$ and the parameters
associated to these functions are the same for all pairs.

The functions $f$ and $g$ are chosen according to the model of Mimura and
Murray \cite{murray_artigo2}:

\begin{equation}
\begin{array}{cc}
  f(u,v)&=\left(\displaystyle\frac{a+bu-u^2}{c}-v\right)u \\
  g(u,v)&=[u-(1+dv)]v, 
\end{array}
\label{eq2rev}
\end{equation}

\noindent where $a$, $b$, $c$ and $d$ are positive parameters that will
be fixed to $a=35$, $b=16$, $c=9$ and $d=0.4$ throughout this paper
\cite{murray_artigo2}.

Both the prey per capita growth rate and the pradator per capita death
rate are density dependent. The hump effect that can be noted in the
prey growth in $f$ represents what in Biology is called the {\it Allee
Effect} \cite{allee1,allee2,begon}, describing a positive correlation
between population density and per capita growth rate in small
populations. The linear function related to the predator per capita death
rate accounts for intraspecific competition in the predator population. 

The possibility of observing Turing patterns for these equations must be
evaluated via linear analysis. Here we show the analysis for the case of a
single prey-predator pair. The general case with $n$ pairs is slightly
more complicated, but can be done following the same lines.

\section{Linear stability analysis}

In this section we briefly review the stability analysis of network
organized systems. For simplicity we consider only one pair of prey and
predator, since the methodology generalizes immediately to the case of
multiple pairs. 

The equilibrium populations in the absence of diffusion,
$(\bar{u},\bar{v})$, are the positive solution of:
\begin{equation}
\begin{array}{cc}
  f(\bar{u},\bar{v})=0  \\
  g(\bar{u},\bar{v})=0
\end{array}
\label{eq:3rev}
\end{equation}

For diffusion driven instability to take place, the equilibrium must be
stable against small perturbations in the absence of diffusion
($\epsilon=0.0$) and go unstable, when diffusion is considered.

Let 
\begin{equation}
	(u_i,v_i)=(\bar{u},\bar{v})+(\delta u_i,\delta v_i)
\label{eq3}
\end{equation}
be small perturbations to the fixed point $(\bar{u},\bar{v})$ at site
$i$. Substituting (\ref{eq3}) in (\ref{eq1}) and linearizing, we obtain
\begin{equation}
\begin{array}{cc}
 \displaystyle\frac{d}{dt}\delta u_i = f_u\delta u_i+f_v\delta
v_i+\varepsilon\displaystyle\sum_{j=1}^N L_{ij}\delta u_i  \\
 \displaystyle\frac{d}{dt}\delta v_i = g_u\delta u_i+g_v\delta
v_i+\sigma\varepsilon\displaystyle\sum_{j=1}^N L_{ij}\delta v_i
\label{eq4}
\end{array}
\end{equation}
where the derivatives are evaluated at the equilibrium.

Since we are dealing with network-organized systems, it is convenient to
expand the perturbations in the basis formed by the eigenvectors of the
Laplacian matrix, \{$\vec{\Phi}^{\alpha}$\} \cite{nakao}. Here
$\alpha=1,\cdots,N$ represent different modes, in direct analogy with the
Fourier modes that appear in continuous systems where the Laplacian is
the usual operator $\nabla^2$. We find
\begin{equation}
\begin{array}{cc}
\delta u_i(t) = \displaystyle\sum_{\alpha=1}^N
c_{\alpha}exp[\lambda_{\alpha}t]\phi_i^{(\alpha)} \\
\delta v_i(t) = \displaystyle\sum_{\alpha=1}^N
c_{\alpha}B_{\alpha}exp[\lambda_{\alpha}t]\phi_i^{(\alpha)} 
\label{eq5}
\end{array}
\end{equation}
	
Substituting (\ref{eq5}) in (\ref{eq4}) and using $\sum_{j=1}^N L_{ij}
\Phi_j^{(\alpha)} = \varLambda_{\alpha}\Phi_i^{(\alpha)}$, we obtain, for
each mode $\alpha$: 
\begin{equation}
\lambda_{\alpha} \left(\begin{array}{c}
     1 \\
     B_{\alpha}
\end{array}\right) =\left(\begin{array}{cc}                               
             f_u+\varepsilon\Lambda_{\alpha} & f_v\\
g_u &	g_v +
\sigma\varepsilon\Lambda_{\alpha}                                         
  \end{array}\right)\left(\begin{array}{c}
  1 \\
  B_{\alpha}     
\end{array}\right)
\label{eq6}
\end{equation}

The matrix obtained in (\ref{eq6}) is the Jacobian of the system
with diffusion. The linear growth rates, $\lambda_{\alpha}$, of
each mode are, as expected, the eigenvalues of the Jacobian matrix.
Turing instability appears when one of the modes becomes
unstable. At the threshold, $Re(\lambda_{\alpha})=0$ for some
$\alpha=\alpha_c$ and $Re(\lambda_{\alpha})<0$ for all other modes.

Above this threshold $Re(\lambda_{\alpha_c})>0$ and perturbations 
grow in time according to $exp[\lambda_{\alpha}t]$, eventually forming
the stationary Turing pattern. A necessary condition for this is that
the solutions of (\ref{eq4}) are confined, otherwise the perturbation
would diverge.

\begin{figure}[ht]
\centering
\includegraphics[height=6cm]{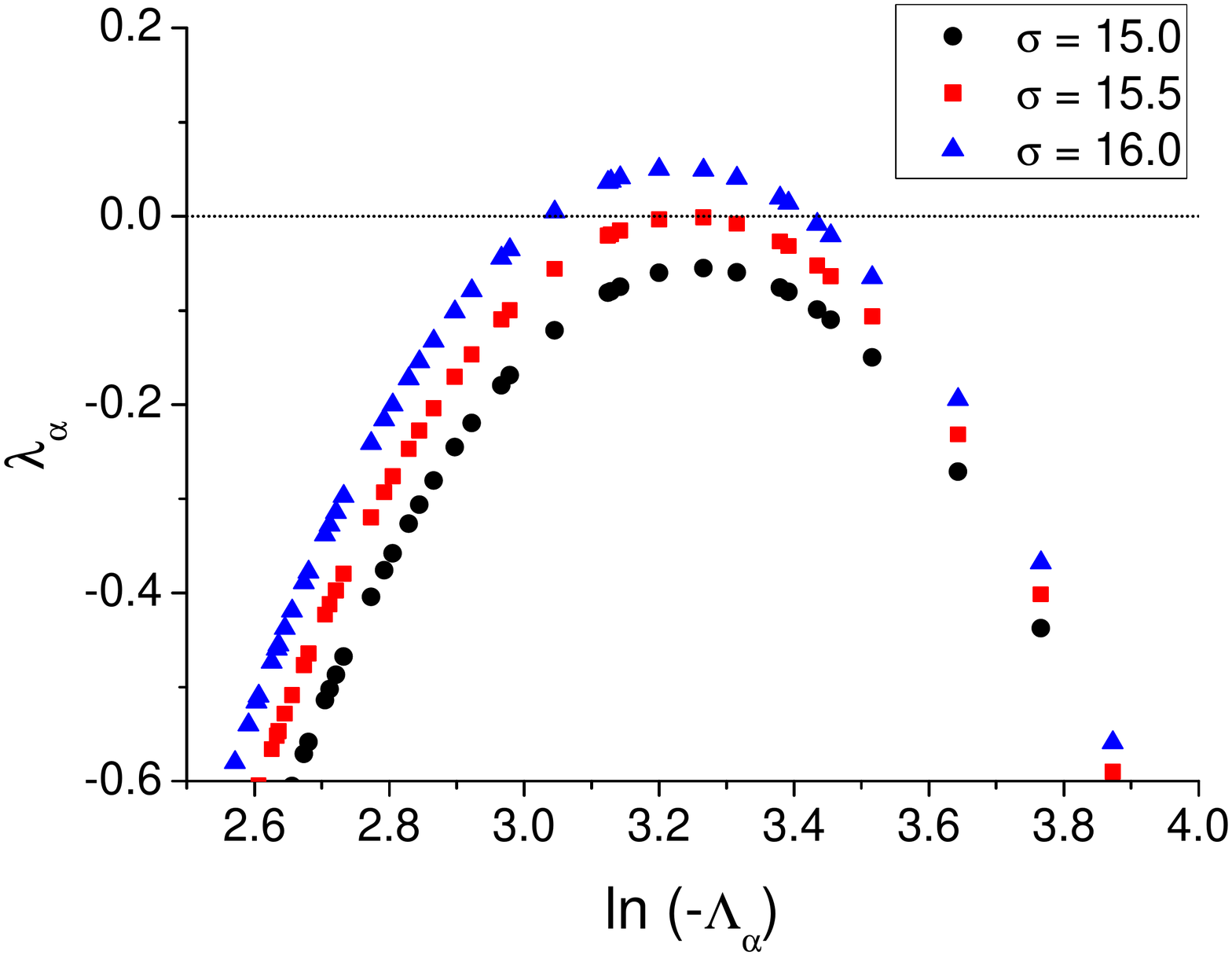}
\caption {(Color online) Linear growth rates, $\lambda_{\alpha}$, as a function
of the eigenvalues of the Laplacian, $\Lambda_{\alpha}$, for a
Barab\'{a}si-Albert network with $N=200$ and $\langle k\rangle=10$. In all
the cases $\epsilon=0.06$ and three different values of $\sigma$ are shown
for comparison. Modes with $\lambda_{\alpha}>0.0$ are observed for $\sigma
> \sigma_c = 15.5$.}
\label{fig2}
\end{figure}

Figure \ref{fig2} shows the linear growth rates, $\lambda_{\alpha}$, 
as a function of the eigenvalues of the Jacobian, $\Lambda_{\alpha}$, 
when we consider the functions (\ref{eq2rev}), with parameters $a=35.0$,
$b=16.0$, $c=9.0$, $d=0.4$ and $\epsilon=0.06$, and a network of $N=200$
nodes with power law degree distribution constructed according to the
Barab\'{a}si-Albert algorithm \cite{revbarabasi}. Below the critical value
$\sigma_c = 15.5$ (see appendix \ref{appa}) $\lambda_{\alpha}<0.0$ for all
the modes and the homogeneous state (\ref{eq:3rev}) is stable.

\section{2 species}

We first review the case of two species as a reference to the more
complex patterns we study in the following sections. We consider a network
with $N=1000$ nodes, constructed according to the Barab\'{a}si-Albert
model \cite{revbarabasi}. The populations of preys, $u_i$, and predators,
$v_i$, defined in each node $i$, interact locally and diffuse through the
network nodes according to the equations (\ref{eq1}), with $l=1$ (and 
$u_i^{(0)}=v_i^{(2)}=0$). Equations (\ref{eq1}) are numerically
integrated until a stationary distribution of the species abundance is
obtained.

\begin{figure}[ht]
\subfigure[]{\includegraphics[height=5.5cm]{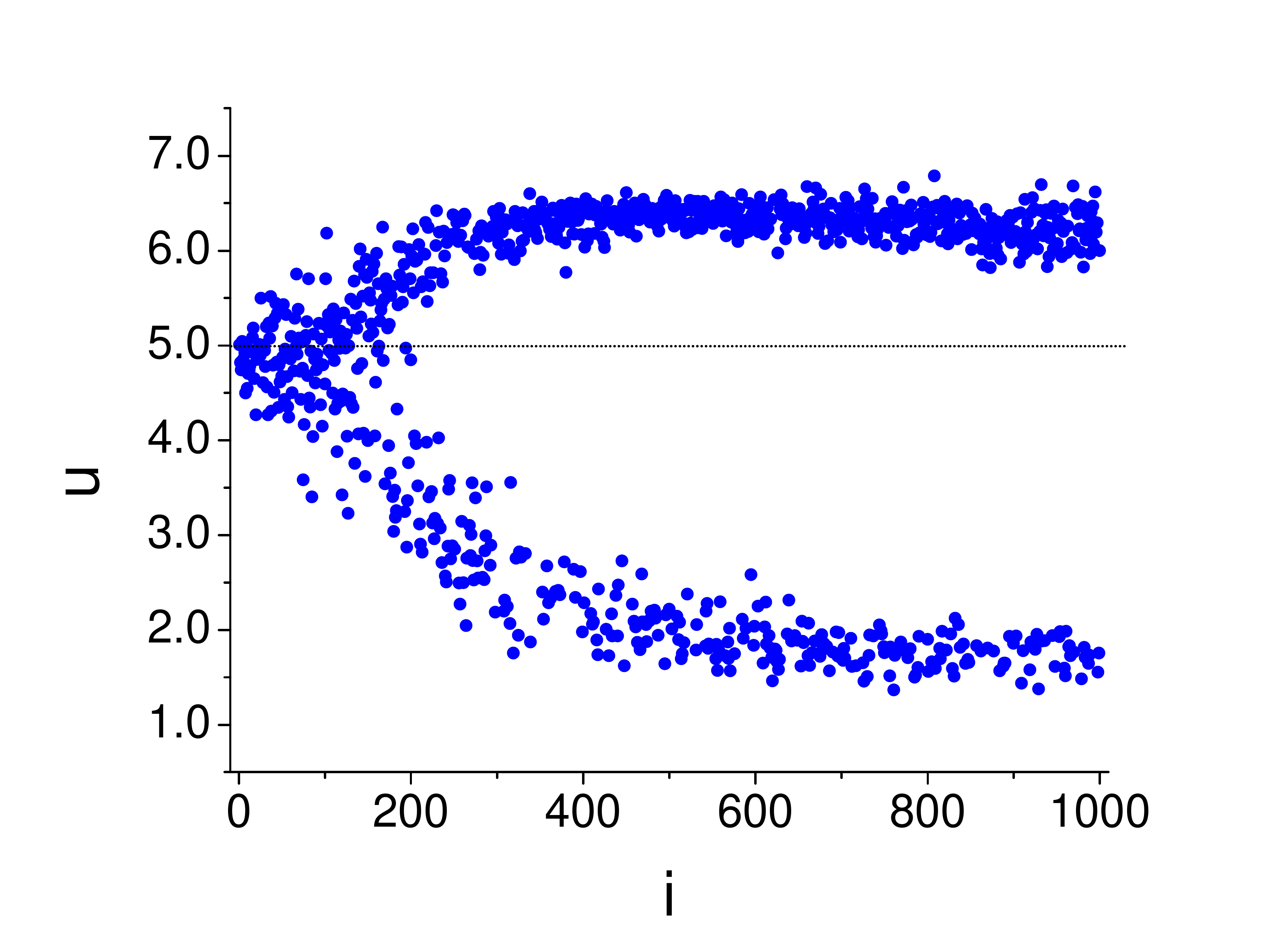}
\label{fig3a}}
\subfigure[]{\includegraphics[height=5.5cm]{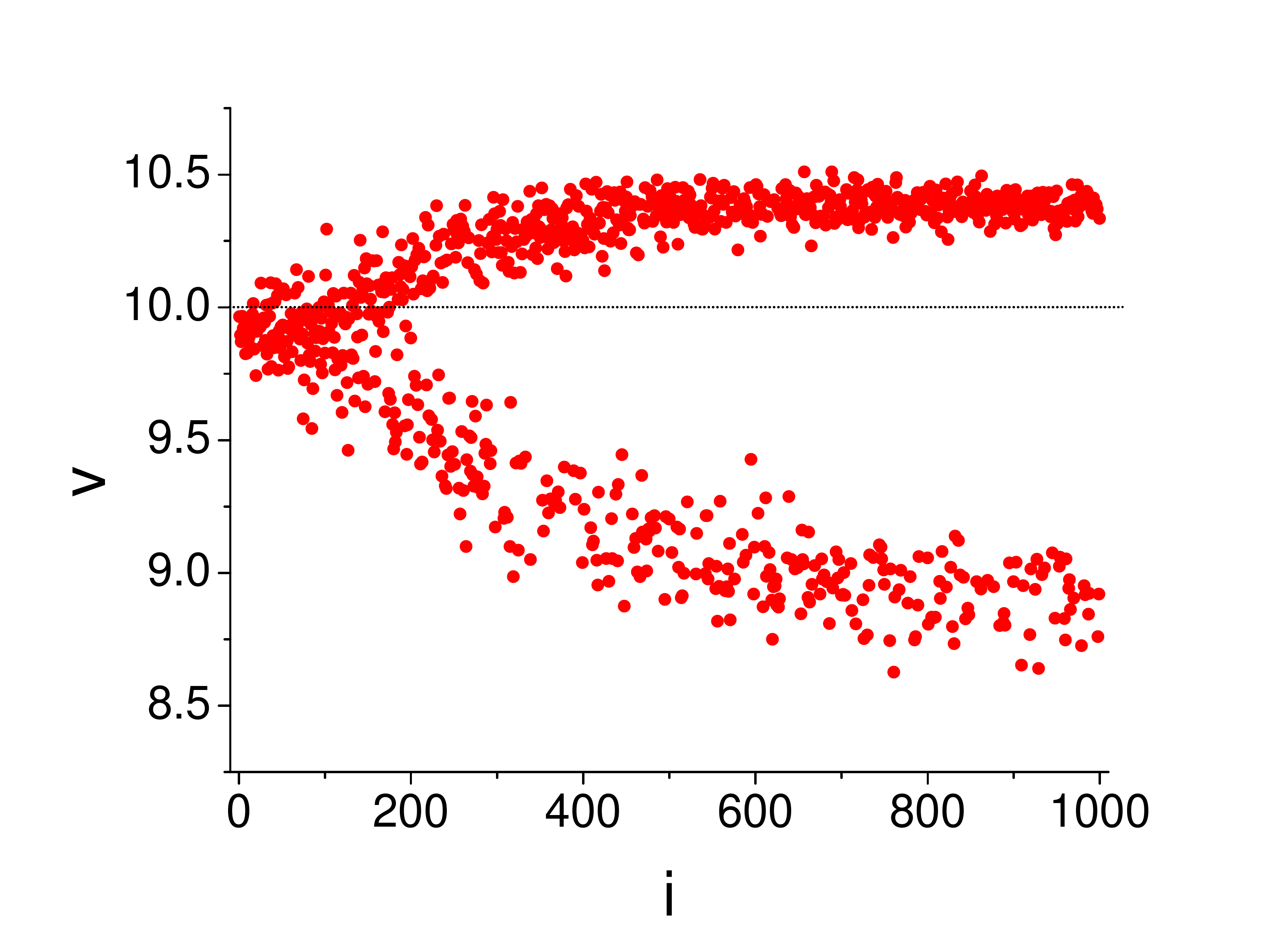}
\label{fig3b}}
\caption {(Color online) Stationary abundance patterns for a single predator-prey pair
as a function of node index $i$ for $\epsilon=0.12$ and $\sigma=20.0$ for
(a) preys and (b) predators. The lines in $u_i=5.0$ and $v_i=10.0$
indicate the values of the homogeneous state, which is a fixed point for
this set of parameters.}
\label{fig3}
\end{figure}

Figure \ref{fig3} shows the stationary abundance patterns of
preys, figure \ref{fig3a}, and predators, figure \ref{fig3b}, as a
function of node index $i$, for $\epsilon=0.12$ and $\sigma=20.0$. The
nodes are ordered according to decreasing degree $k_i$. 

The pattern of prey distribution is formed by two groups of nodes 
presenting significant differentiation in relation to the homogeneous
state: a group with high abundance (values of $u_i$ well above $\bar{u}$)
and a group with low abundance (values of $u_i$ well below $\bar{u}$). The
pattern of predators follows directly the pattern of the preys: nodes
with large abundance of preys ($u_i>\bar{u}$) also have large abundance of
predators ($v_i>\bar{v}$) and vice versa.

\section{4 species}

The 4 species system is described by Eq.(\ref{eq1}) with $l=1,2$
(and $u_i^{(0)}=v_i^{(3)}=0$). The equations have a homogeneous
equilibrium point that depends on the coupling parameter $\gamma$, as
displayed by the table \ref{table1}. The populations of preys and
predators decrease as $\gamma$ increases.

\begin{table}[ht]
\centering 
\begin{tabular}{|c|cccc|} 
\hline
$\gamma$  & $u^{(1)}$ & $v^{(1)}$ & $u^{(2)}$ & $v^{(2)}$ \\ [0.5ex]
\hline
0.002 & 4.989 & 9.973 & 4.993 & 9.995 \\ 
0.01  & 4.945 & 9.863 & 4.966 & 9.977 \\
0.05  & 4.726 & 9.314 & 4.837 & 9.889 \\ 
\hline 
\end{tabular}
\caption{Homogeneous fixed points for different values of $\gamma$ for
the four species system.}
\label{table1}
\end{table}

\begin{figure}[ht]
\includegraphics[height=6cm]{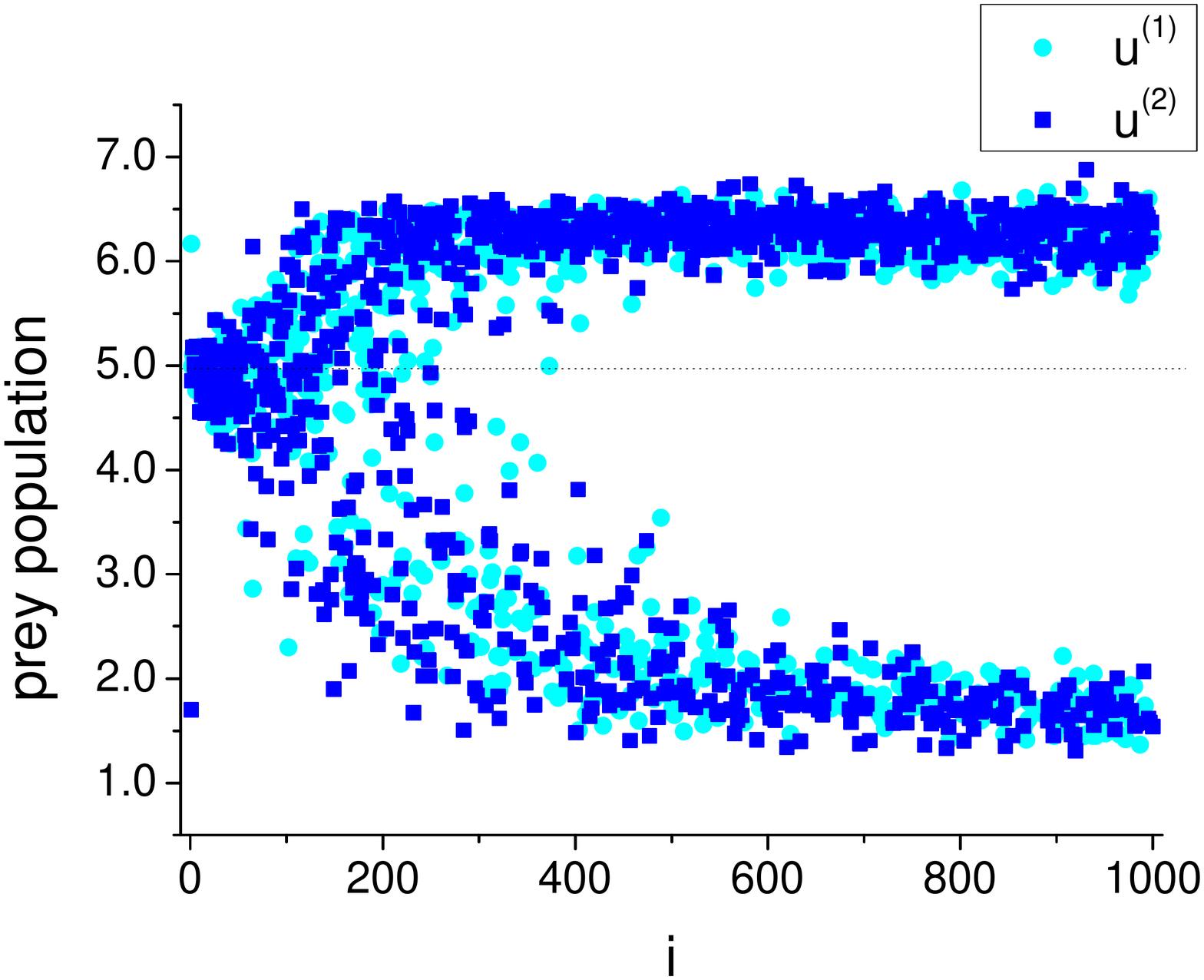}
\caption {(Color online) Stationary abundance patterns for $u^{(1)}$ and $u^{(2)}$ as a
function of node index $i$ for $\epsilon=0.12$, $\sigma=20.0$, $\phi =0.5$
and $\gamma=0.002$.}
\label{fig4}
\end{figure}

The stationary patterns of preys $u^{(1)}$ and $u^{(2)}$ as a function of
the node index $i$ are shown in figure \ref{fig4}. These patterns of
abundance (and also those of $v^{(1)}$ and $v^{(2)}$) are not very
different from each other or from the previous case shown in figure
\ref{fig3}. In particular, both types of preys and predators present the
separation of nodes in high abundance and low abundance groups.

\begin{figure}[ht]
\subfigure[]{\includegraphics[height=3.7cm]{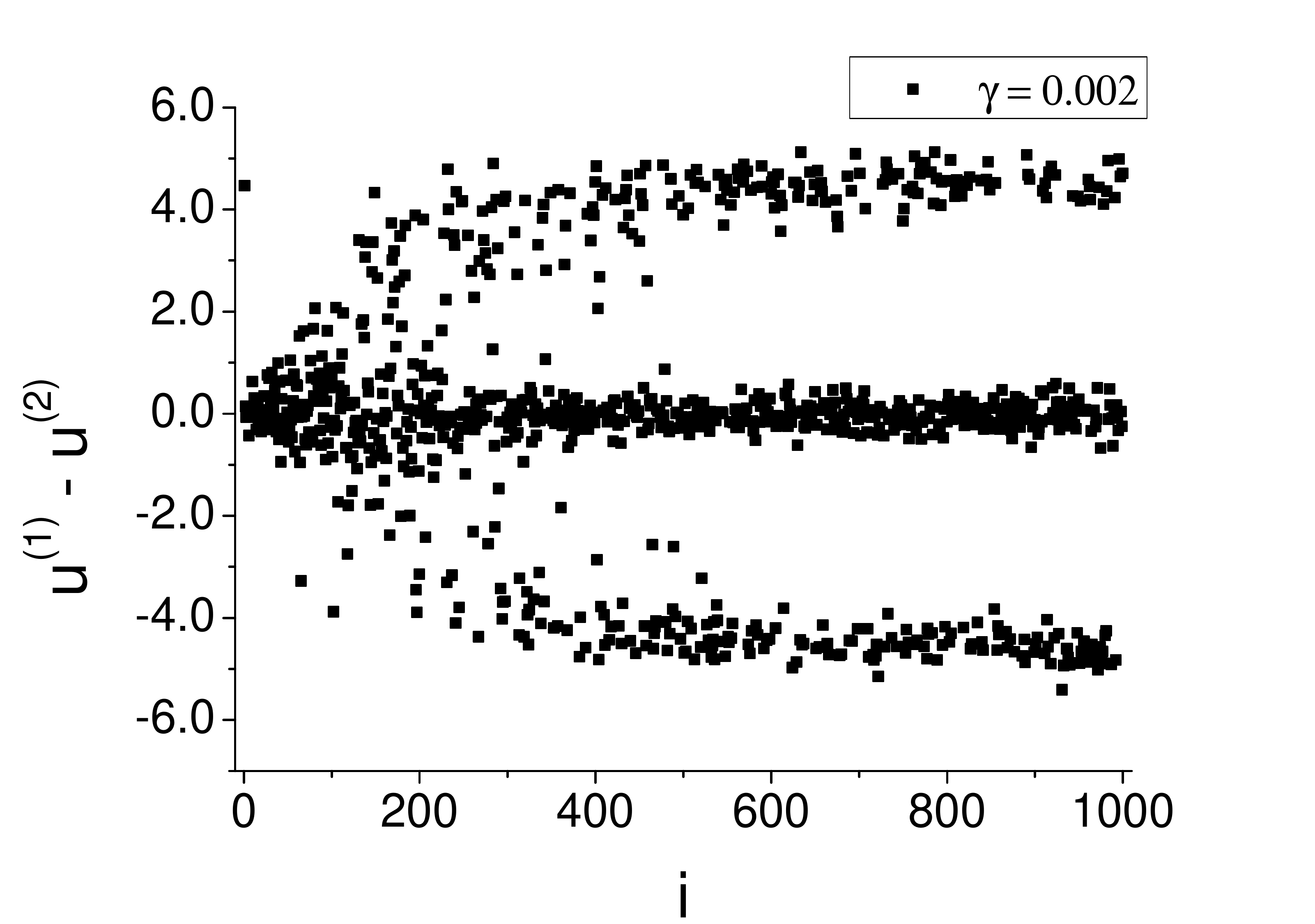}
\label{fig5a}}
\subfigure[]{\includegraphics[height=3.7cm]{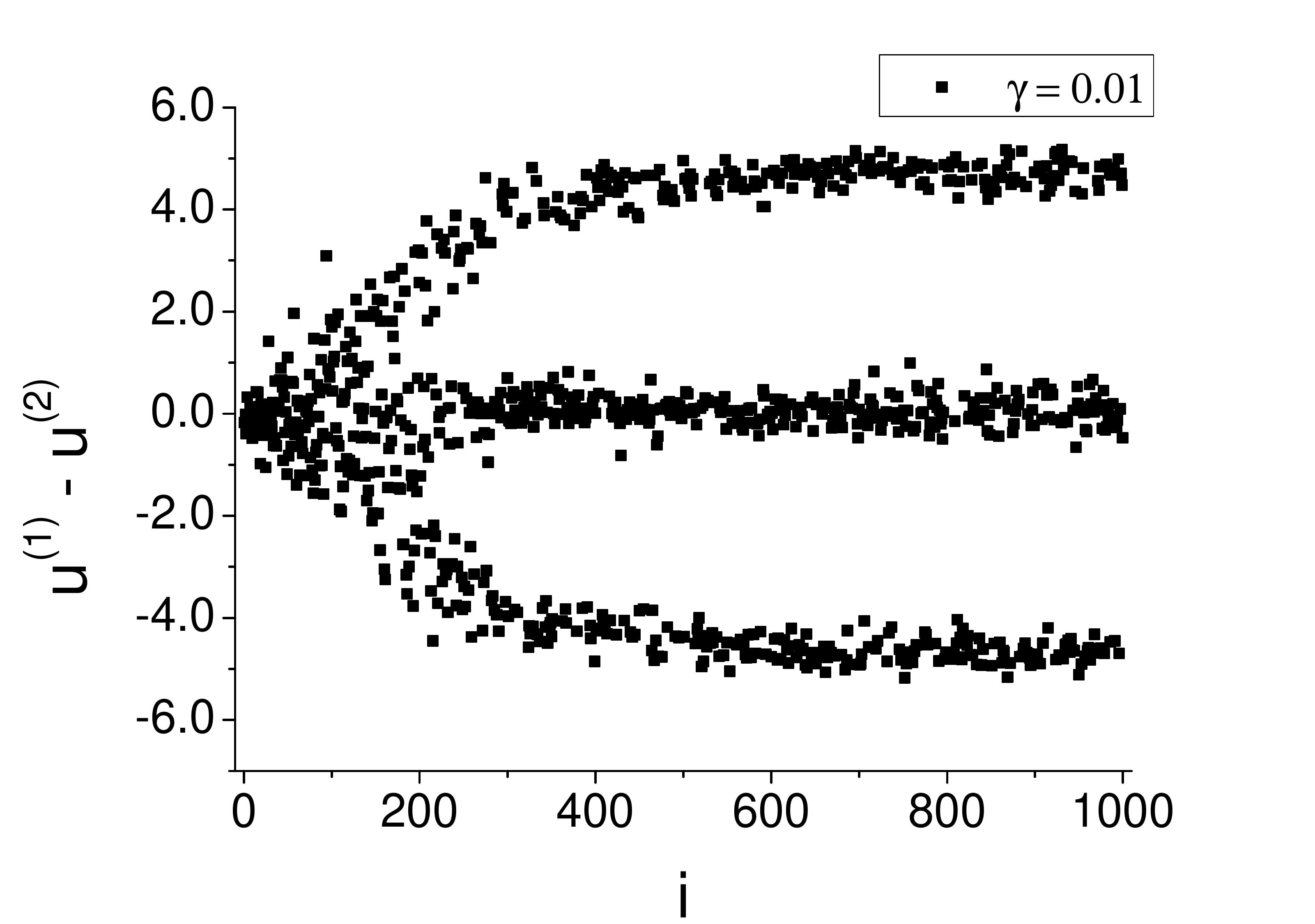}
\label{fig5b}}
\subfigure[]{\includegraphics[height=3.7cm]{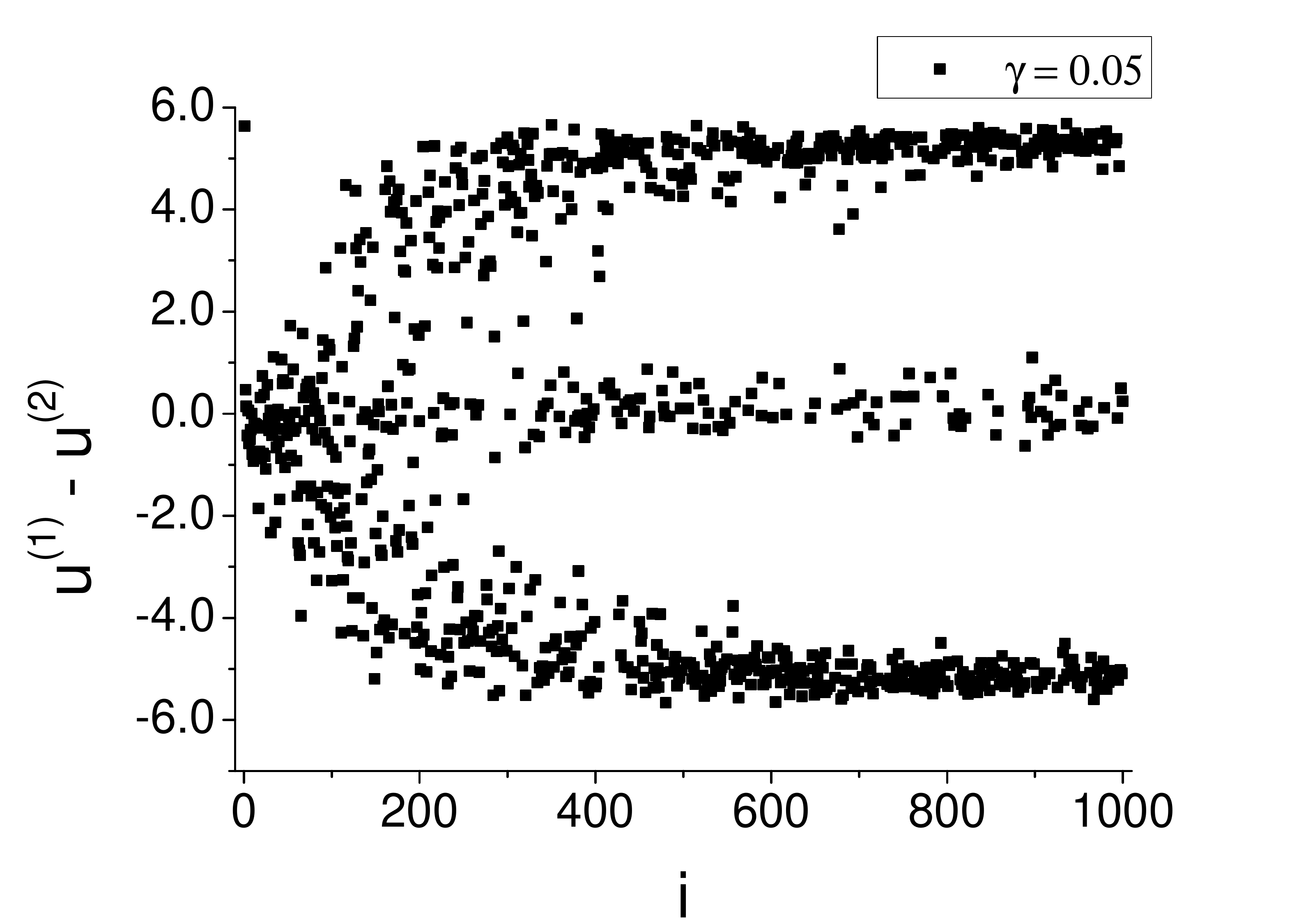}
\label{fig5c}}
\caption {Stationary patterns for the difference $u^{(1)}-u^{(2)}$ as a
function of node index $i$ for (a) $\gamma=0.002$, (b) $\gamma=0.01$ 
and (c) $\gamma=0.05$. In all cases $\epsilon=0.12$, $\sigma=20.0$ and
$\phi =0.5$.}
\label{fig5}
\end{figure}

However, this similarity is partly an illusion, having to do with the way
the data is plotted. Indeed, a new underlying pattern arises when
difference between the prey abundances $u_i^{(1)}-u_i^{(2)}$ is
plotted, as shown in figure \ref{fig5} for different values of the
coupling strength $\gamma$.

In all cases it is possible to distinguish three main branches: the
upper branch, where $u^{(1)}-u^{(2)} \approx 4$, corresponding to
nodes where $u^{(1)}$ is abundant but $u^{(2)}$ is not; the lower branch,
where $u^{(1)}-u^{(2)} \approx -4$ where the abundances are reversed; and
the middle branch, where $u^{(1)}-u^{(2)} \approx 0$ and $u^{(1)}$ and
$u^{(2)}$ have similar abundances.  This configuration of branches 
can be derived via a mean field approximation \cite{meanf1,meanf2}, as
discussed in appendix \ref{appb} and displayed in Fig. \ref{fig6}.

\begin{figure}[ht]
\includegraphics[height=6cm]{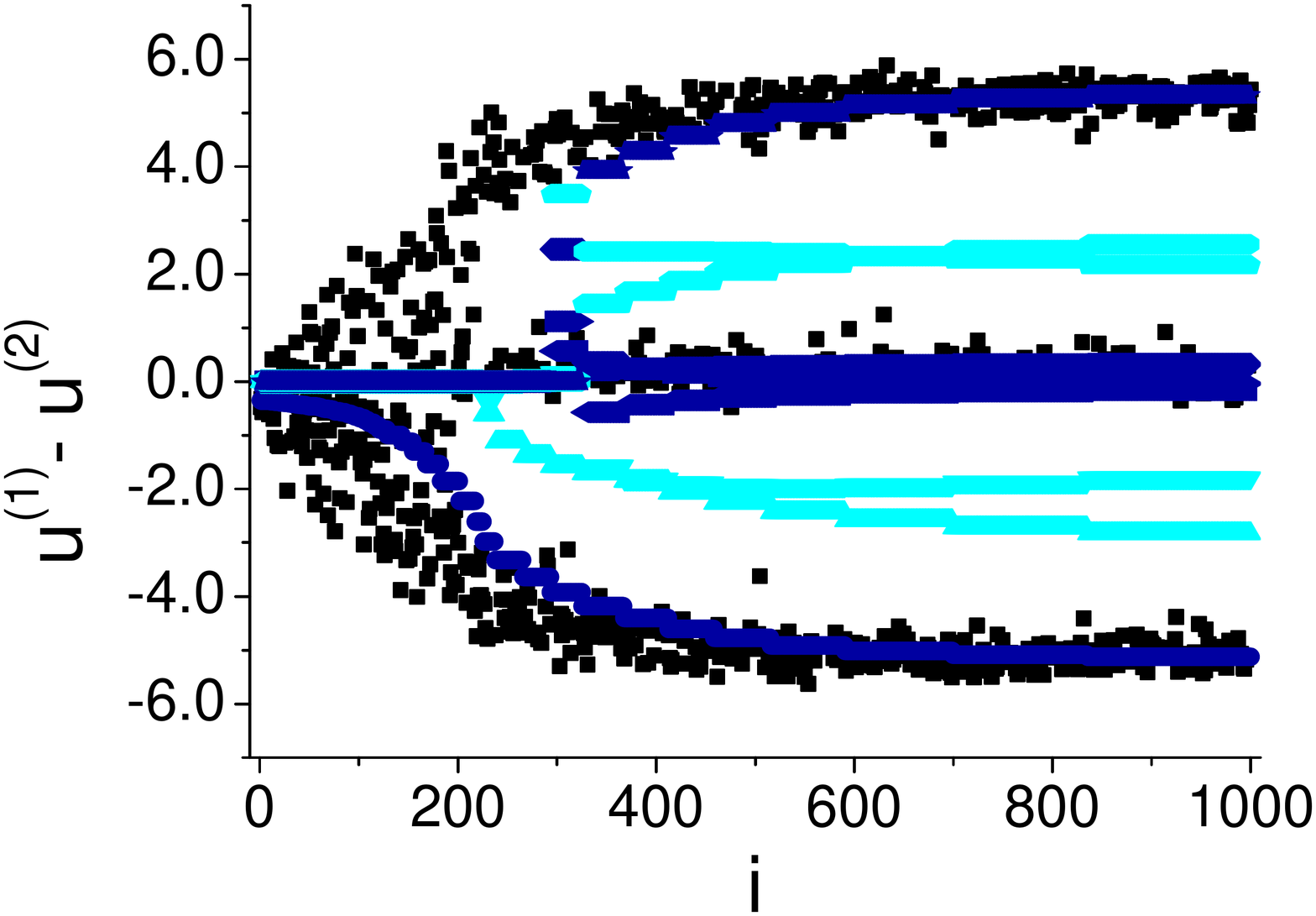}
\caption {(Color online) Stationary pattern for the difference $u^{(1)}-u^{(2)}$ as
obtained from simulations (black squares) and from the mean field
approximation (blue) for $\epsilon=0.12$, $\sigma=20.0$, $\phi =0.5$ and
$\gamma=0.05$. The cyan lines show unstable branches.}
\label{fig6}
\end{figure}

As $\gamma$ increases the middle branch gets less populated and the
nodes are dominated mostly by a single species of prey and predator.
This corresponds to a strong effect of apparent competition driven by
Turing instabilities. The more important is the secondary predation (which
is kept weaker than the direct predation in the each pair), the stronger
is the effect. 

\section{6 species}

\begin{figure}[ht]
\includegraphics[height=6cm]{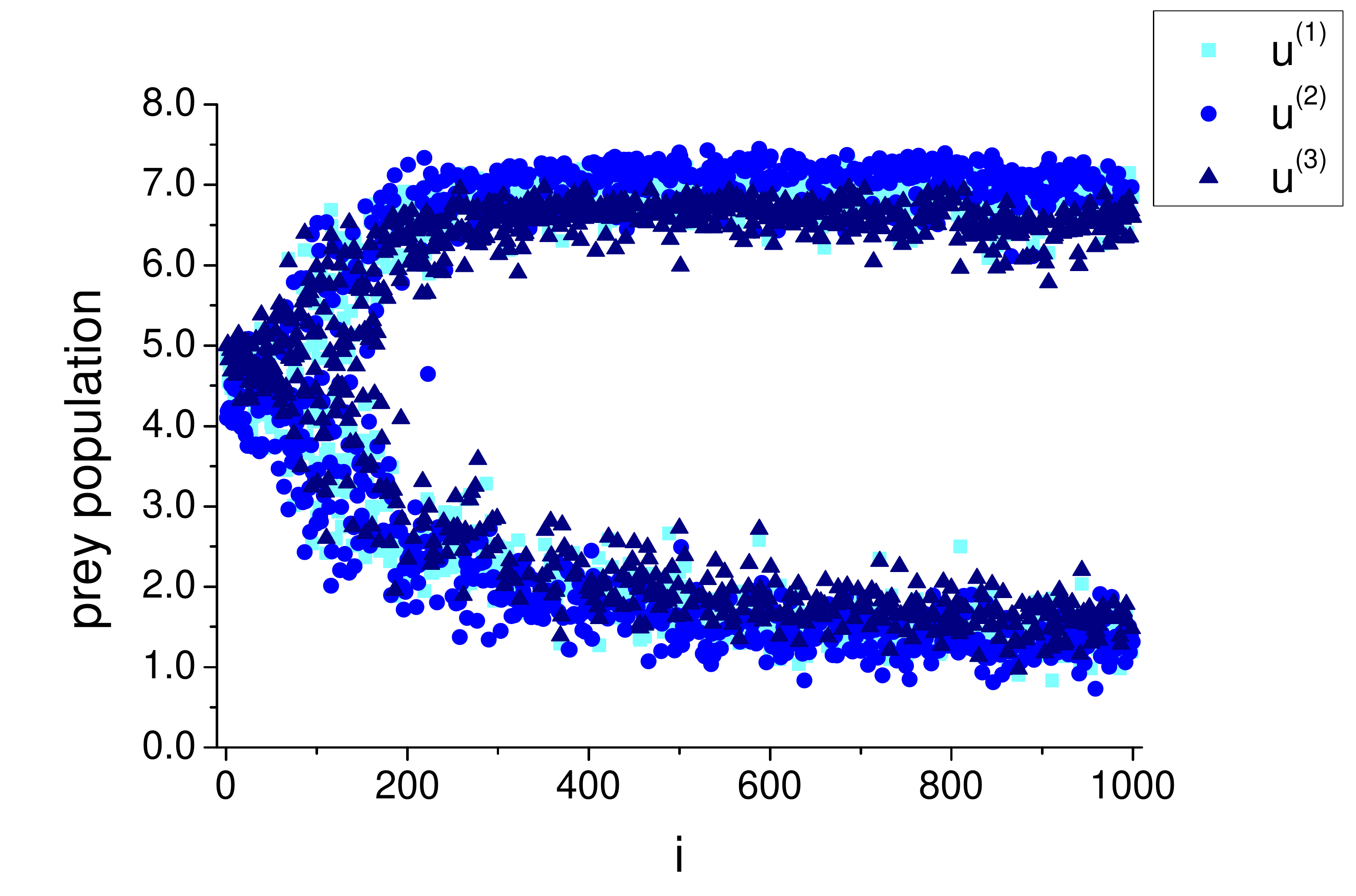}
\caption {(Color online) Stationary patterns of preys distributions as a function
of node index $i$ for the case of three pairs for $\epsilon=0.12$,
$\sigma=20.0$, $\phi =0.5$ and $\gamma=0.05$.}
\label{fig7}
\end{figure}

To investigate if the negative correlation between preys of coupled
pairs also occur in larger trophic chains we consider a system with 6
species, again given by equation (\ref{eq1}) with $l=1,2,3$ (and
$u_i^{(0)}=v_i^{(4)}=0$). The stationary patterns of preys 
distributions are shown in figure \ref{fig7}.

Once again, for each prey species, the nodes cluster into groups of high
and low abundances. The analysis of the correlations between different
prey species, however, is now more involved. We first define the quantity:
\begin{equation}
\sigma_i^{(l)} = sng(u_i^{(l)}-\bar{u}^{(l)}) = \left\{
\begin{array}{rcl}
+1,& \mbox{if}\ \ u_i^{(l)}>\bar{u}^{(l)}\\
-1, & \mbox{if}\ \ u_i^{(l)}<\bar{u}^{(l)}
\end{array}
\right.,
\label{eq48}
\end{equation}
where $\sigma_i^{(l)}$ indicates if the $l$-th prey population at node $i$
has high ($\sigma_i^{(l)}=+1$) or low ($\sigma_i^{(l)}=-1$) abundance with
respect to the homogeneous value. 

\begin{figure*}[!t]
\subfigure[]{\includegraphics[height=5cm]{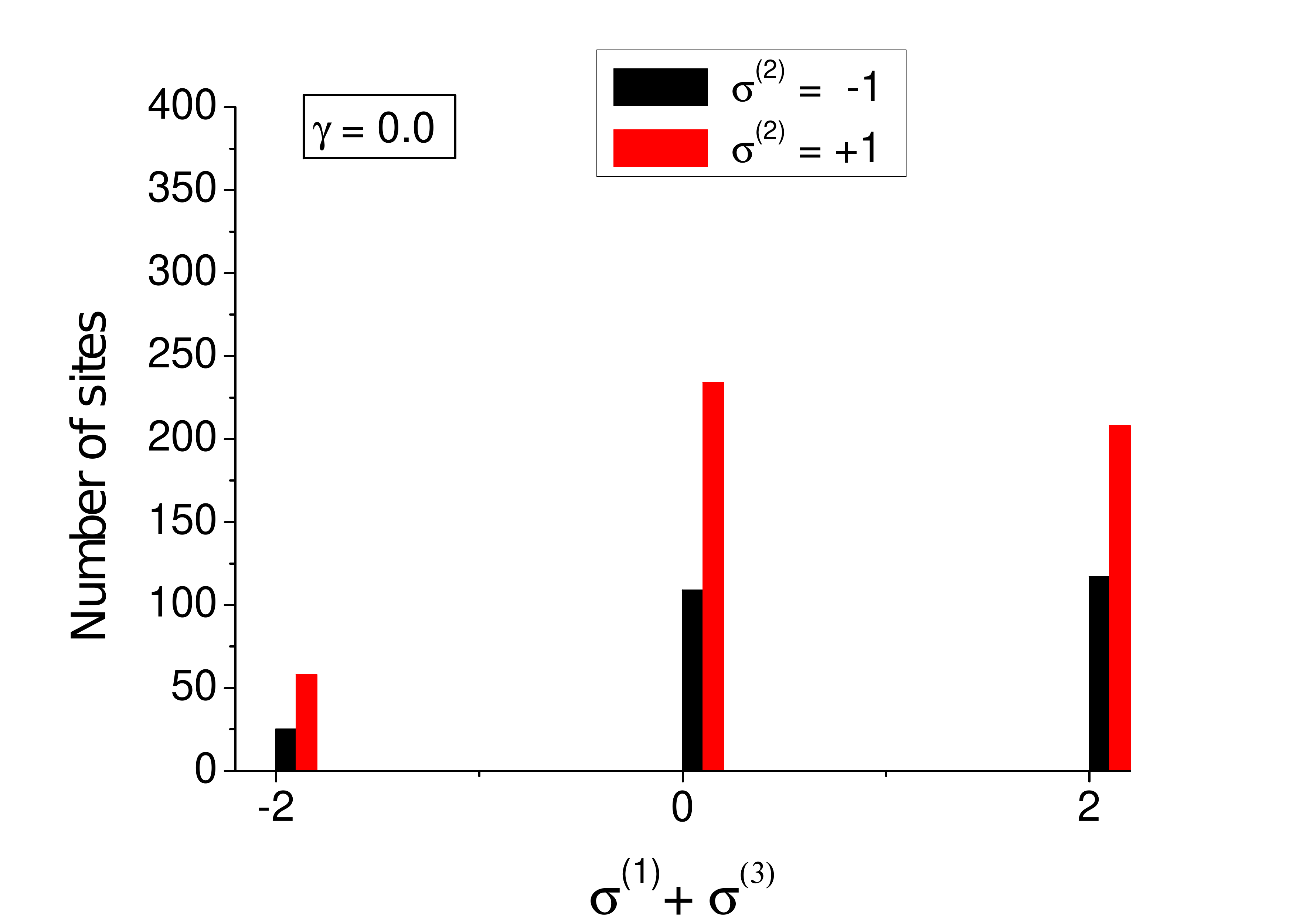}
\label{fig8a}}
\hspace{-0.5cm}
\subfigure[]{\includegraphics[height=5cm]{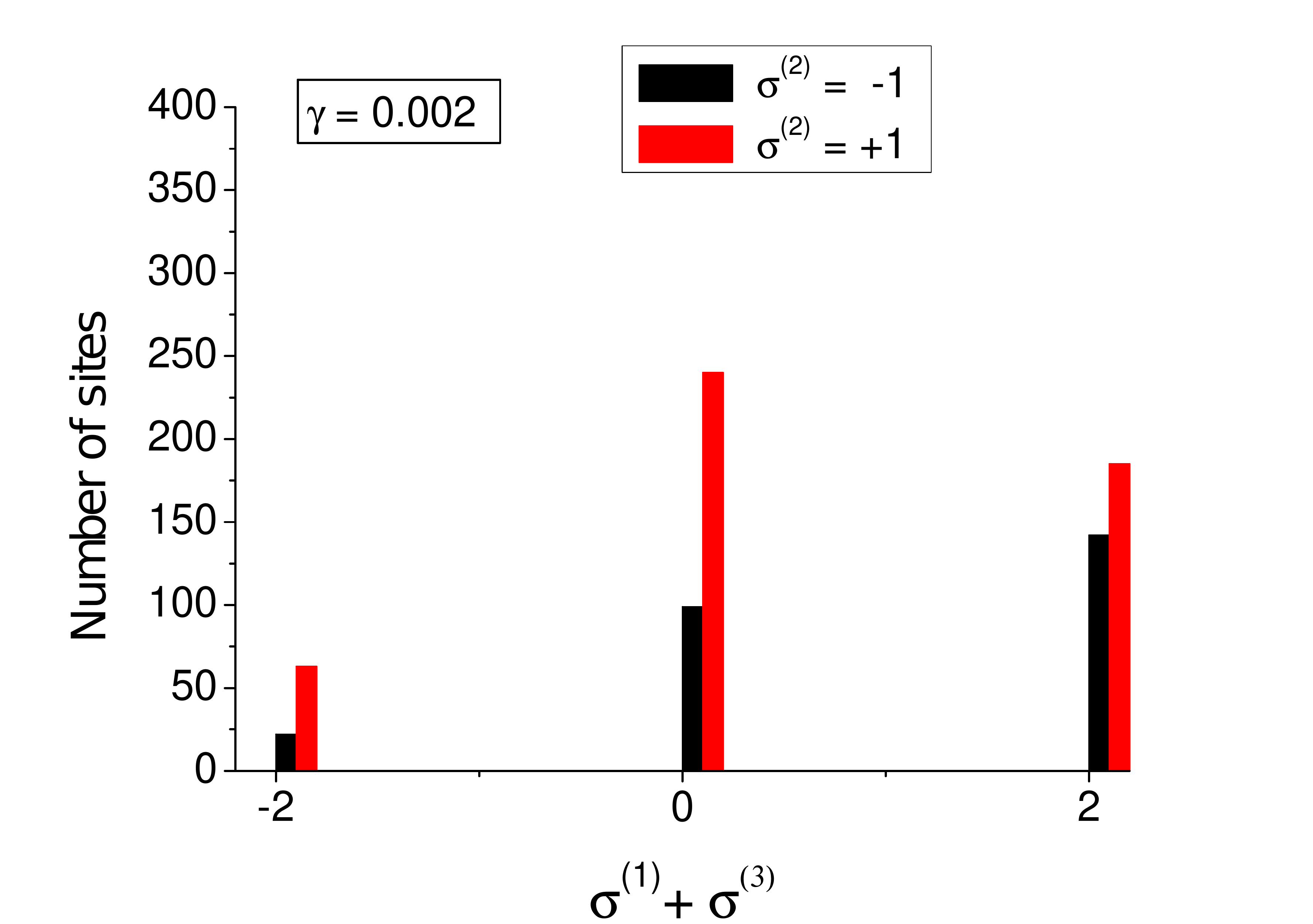}
\label{fig8b}}
\hspace{-0.5cm}
\subfigure[]{\includegraphics[height=5cm]{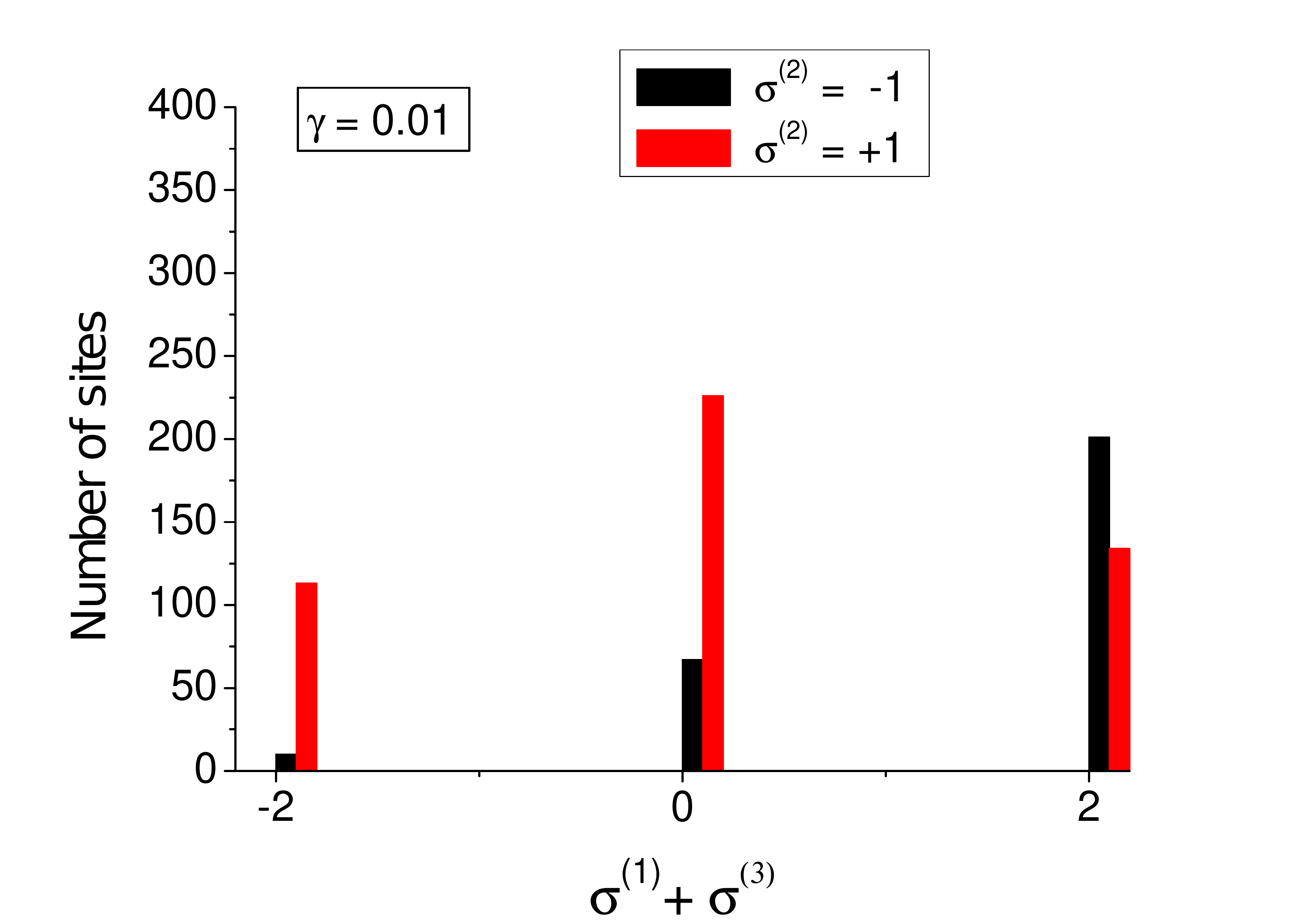}
\label{fig8c}}
\hspace{-0.5cm}
\subfigure[]{\includegraphics[height=5cm]{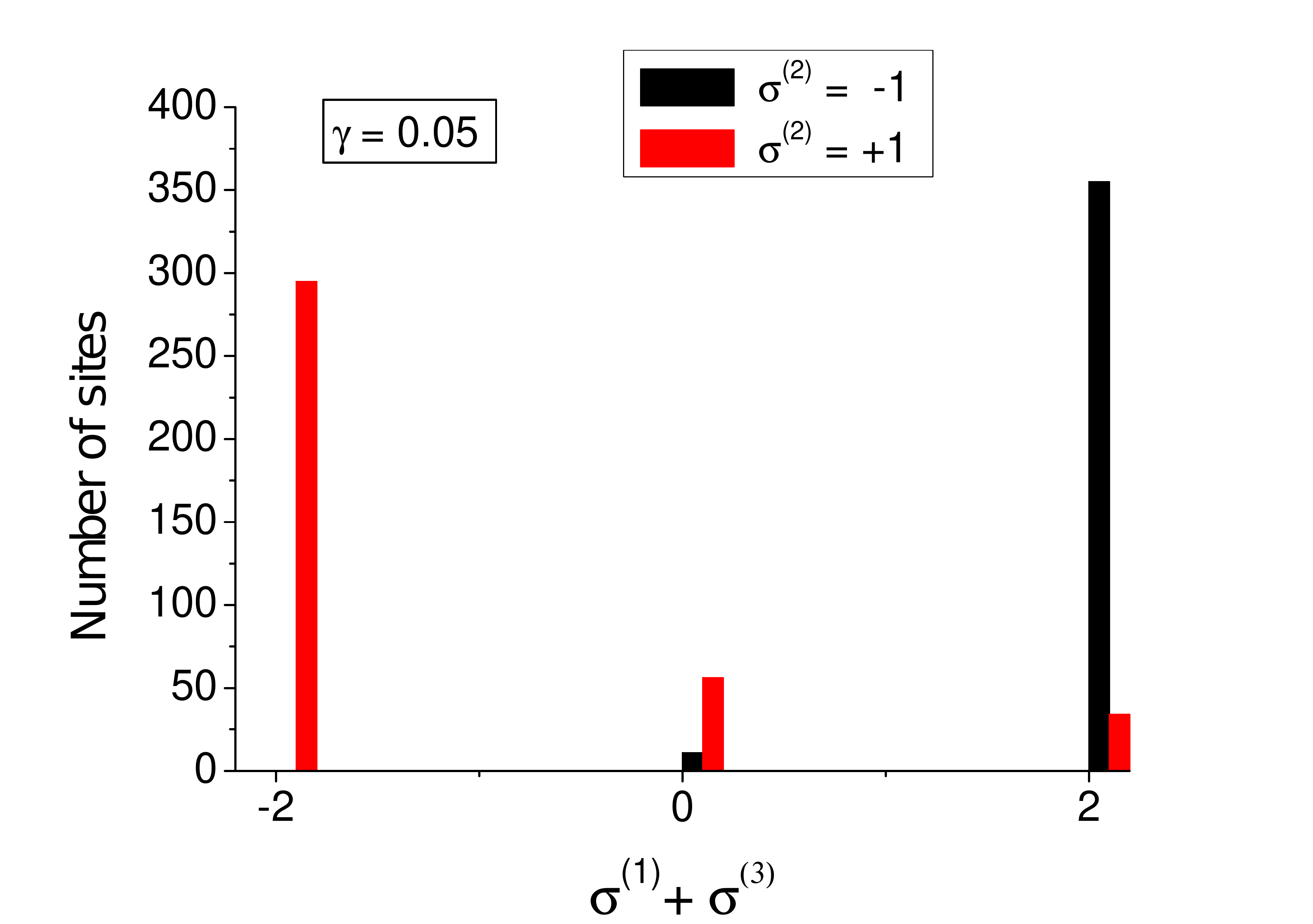}
\label{fig8d}}
\caption {(Color online) Histogram of nodes with different values of
$\sigma^{(1)}+\sigma^{(3)}$ for $\sigma^{(2)}=-1$ (black bars) and
$\sigma^{(2)}=+1$ (red bars) for: (a) $\gamma=0.0$, (b) $\gamma=0.002$,
(c) $\gamma=0.01$ and (d) $\gamma=0.05$} 
\label{fig8}
\end{figure*}

Second, we separate the nodes in two groups: those with $\sigma^{(2)}=+1$
and those with $\sigma^{(2)}=-1$. Since nodes with large $k_i$ are not
sensitive to the coupling, we restrict this analysis to nodes with $i \geq
250$, for which the differentiation is more evident. Finally we focus on
the value of the sum $\sigma^{(1)}+\sigma^{(3)}$ for these nodes. The
three possible values of this sum indicate the following situations: if
$\sigma^{(1)}+\sigma^{(3)}=+2$, both $u^{(1)}$ and $u^{(3)}$ have high
abundance in the node; if $\sigma^{(1)}+\sigma^{(3)}=-2$, both $u^{(1)}$
and $u^{(3)}$ have low abundance and if $\sigma^{(1)}+\sigma^{(3)}=0$,
$u^{(1)}$ and $u^{(3)}$ have opposed abundance characteristics. If the
hypothesis of negative correlation is to be valid, the group of nodes with
$\sigma^{(2)}=+1$ must have most of its node with
$\sigma^{(1)}+\sigma^{(3)}=-2$ and the group with $\sigma^{(2)}=-1$ must
have most of its nodes with $\sigma^{(1)}+\sigma^{(3)}=+2$. The results
are shown in figure \ref{fig8} in the form of histograms. 

In Fig. \ref{fig8a} $\gamma=0.0$ and the three preys distributions are
uncorrelated. Figures \ref{fig8b} and \ref{fig8c} display cases with
increasing values of $\gamma$. As the coupling strength increases, the
number of nodes with $\sigma^{(1)}+\sigma^{(3)}=+2$ increases in the group
with $\sigma^{(2)}=-1$ and similarly with the number of nodes with
$\sigma^{(1)}+\sigma^{(3)}=-2$ in the group where $\sigma^{(2)}=+1$. This
separation is evident in figure \ref{fig8d}, where $\gamma=0.05$, where
it is clear that most of the nodes where $u^{(2)}$ has large abundance
display low abundances of both $u^{(1)}$ and $u^{(3)}$ and
vice versa, showing the persistence of the negative correlation between
preys of coupled pairs. 

\section{Discussion}

We have studied the formation of Turing patterns in an extended
prey-predator system, considering trophic chains composed of 1, 2 and 3
prey-predator pairs, coupled by cross predation and 
dispersing through the connected nodes of a complex network. We detected
the emergence of negative correlations between the populations of preys of
coupled pairs in each node, even though there are no direct competition
between preys in the equations. This effect, known in Biology as
apparent competition \cite{appcomp1,appcomp2}, is triggered here by the
Turing instabilities, and not by the local interactions. 

The description of fragmented landscapes as complex networks is
relatively recent in ecology \cite{fragment}. Although large landscape networks 
have been studied \cite{minor}, most of the empirical work
has dealt with a relatively small number of patches \cite{urban} and it
is not obvious that the patterns observed here for networks with $N=1000$
nodes persist in smaller sets. We have checked that for $N$ as low as
$100$ the same pattern of apparent competition can be clearly identified,
but not so much for $N=50$, which seems to be a limiting size for the
present set of parameters.

Another important concern in the application of our results to realist
ecological problems is the topology of the network. All numerical
simulations presented in the previous sections were performed for networks
exhibiting power law decay of the degree distribution, that results from
the application of the Barab\'{a}si-Albert algorithm. Natural landscape
networks can exhibit significant heterogeneity in the degree distribution
\cite{fortuna}, but are not necessarily scale free. In order to
verify the robustness of our results against changes in the network
topology we have also simulated networks with Poisson degree distribution,
associated to random networks. We found that the negative correlations
between preys still holds for $N=1000$ and average degree $\langle k
\rangle = 20$.

The occurrence of Turing patterns in real ecological systems is still an
open question. This is in part due to the difficulties in conducting
controlled ecological field experiments to distinguish between
patterns related to space heterogeneity or to intrinsic mechanisms of the
interaction. However, there is growing evidence of species distribution
patterns formed by the Turing mechanism \cite{rietker,maron}. 

Our results point to the possibility that, at least in part, species
abundance patterns might be related to Turing instabilities and not to
environmental heterogeneity. Moreover, strong effects of apparent
competition might emerge spontaneously as Turing patterns, resulting from
diffusion instabilities and not necessarily from local interactions.

\section{Aknowledgements}

It is a pleasure to thank Carolina Reigada for important discussions. This work was partly supported by FAPESP  and CNPq.

\begin{appendix}

\section{Critical value for Turing instability}
\label{appa}

The eigenvalues of the Jacobian matrix of \ref{eq6} are given by the roots
of the characteristic polinomial

\begin{equation*}
\lambda_{\alpha}^2 - \lambda_{\alpha}(f_u + g_v + 
(1+\sigma)\varepsilon\varLambda_{\alpha}) + (f_u +
\varepsilon\varLambda_{\alpha})(g_v +
\sigma\varepsilon\varLambda_{\alpha}) - f_vg_u = 0
\end{equation*}
which are given by

\begin{equation}
\lambda_{\alpha} = \frac{f_u+g_v+(1+\sigma)\varepsilon\varLambda_{\alpha}
\pm
\sqrt{4f_vg_u+(f_u-g_v+(1-\sigma)\varepsilon\varLambda_{\alpha})^2}}{2}.
\label{eq_lambda}
\end{equation}

For each mode $\alpha$ there are two possible values for
$\lambda_{\alpha}$, but only the one associated to the plus sign can
become positive, so we only need to consider this eigenvalue. Solving
$d(\lambda_{\alpha})/d(\varLambda_{\alpha})=0$ we obtain the critical
Laplacian eigenvalue. Substituting this value in \ref{eq_lambda} and
imposing that $Re(\lambda_{\alpha_c})=0$ in the instability threshold, we
obtain the critical value  $\sigma_c$:

\begin{equation}
\sigma_c = \displaystyle\frac{f_ug_v -2f_vg_u+
2\sqrt{f_vg_u(f_vg_u-f_ug_v)}}{f_u^2}.
\end{equation}

\section{Mean field approximation}
\label{appb}

The mean field approximation consists in averaging the heterogeneous
degree distribution of the network by adjusting the strength by which each
node senses the presence of its neighbors. Introducing the local fields

\begin{equation}
	\begin{array}{cc}
		x_i^{l}=\sum_{j=1}^N A_{ij}u_j^{(l)} \\
		y_i^{l}=\sum_{j=1}^N A_{ij}v_j^{(l)}
		\label{loc_field}
	\end{array}
\end{equation}
and substituting in Eq.(\ref{eq1}), we obtain

\begin{align}
	\frac{d}{dt}u_i^{(l)}(t)&=f(u_i^{(l)},v_i^{(l)})-
	\gamma u_i^{(l)}v_i^{(l+1)}+\varepsilon\left(x_i^{l} -
	k_iu_i^{(l)}\right) \notag \\
	\frac{d}{dt}v_i^{(l)}(t)&=g(u_i^{(l)},v_i^{(l)})+
	 \phi\gamma u_i^{(l-1)}v_i^{(l)}
	+\sigma\varepsilon\left(y_i^{l} - k_iv_i^{(l)}\right). 
	\label{loc_field_eq}
\end{align}

We then consider the approximations $x_i^{l} \simeq k_i X^{l}$
and $y_i^{l} \simeq k_i Y^{l}$, where the global fields $X$ and $Y$
are defined as the weighted averages

\begin{equation}
	\begin{array}{cc}
		X^{l}=\sum_{j=1}^N w_j u_i^{(l)} \\
		Y^{l}=\sum_{j=1}^N w_j v_i^{(l)}
		\label{glob_field}
	\end{array}
\end{equation}

\noindent with the weights

\begin{equation}
	w_j = k_j \big/ \sum_{l=1}^N k_l.
\end{equation}

\noindent This choice gives hubs have a stronger influence in the
calculation of the global fields.

With this approximation, and introducing the parameter $\beta(i) =
\varepsilon k_i$, the dynamical system may be written as:

\begin{align}
	\frac{d}{dt}u^{(l)}(t)&=f(u^{(l)},v^{(l)})-
	\gamma u^{(l)}v^{(l+1)}+\beta\left(X^{l} - u^{(l)}\right)
	\notag \\
	\frac{d}{dt}v^{(l)}(t)&=g(u^{(l)},v^{(l)})+
	\phi\gamma u^{(l-1)}v^{(l)}+\sigma\beta\left(Y^{l} -
	v^{(l)}\right), 
	\label{loc_field_eq2}
\end{align}

\noindent where each dynamical variable interacts only with its
associated global field. Since every node now possesses the same dynamical
equation, we may drop the index $i$.

In order to describe the patterns for the difference of prey populations, 
in the case with 2 prey-predator pairs, we define the new variables:
\begin{equation}
\begin{array}{cccc}
	u_\pm = u^{(1)} \pm u^{(2)}\\
	v_\pm = v^{(1)} \pm v^{(2)}
\end{array}
\label{new_var}
\end{equation}
	
The system of equations (\ref{loc_field_eq2}), written with the new
variables (\ref{new_var}), is given by:

\begin{align}    
    \frac{d u_\pm}{dt}&=F_\pm(u_-,u_+,v_-,v_+) - \frac{\gamma}{4}
    (u_-+u_+)(v_+-v_-)+\beta\left(X^{\pm} - u_+\right) \notag \\
    \frac{d v_\pm}{dt}&=G_\pm(u_-,u_+,v_-,v_+) \pm \frac{\phi\gamma}{4}
    (u_-+u_+)(v_+-v_-)+\sigma\beta\left(Y^{\pm} - v_+\right),	
\label{new_var_eq}
\end{align}

\noindent where 
\begin{equation}
\begin{array}{cccc}

F_\pm =f(u^{(1)}(u_-,u_+),v^{(1)}(v_-,v_+))
\pm f(u^{(2)}(u_-,u_+),v^{(2)}(v_-,v_+))\\
		
G_\pm =g(u^{(1)}(u_-,u_+),v^{(1)}(v_-,v_+)) \pm
g(u^{(2)}(u_-, u_+),v^{(2)}(v_-,v_+))
\end{array}
\label{new_din}
\end{equation}
	
\noindent and
\begin{equation}
\begin{array}{cccc}
	X^{\pm}=X^{1} \pm X^{2}\\
	Y^{\pm}=Y^{1} \pm Y^{2}\\
\end{array}
\label{new_fields}
\end{equation}

If the global fields for each dynamical variable are given, the parameter 
$\beta$ may be seen as a bifurcation parameter. It is possible to note a
saddle node bifurcation in the system, and the appearance of new stable
fixed points, when the value of $\beta$ is increased from $\beta=0$.

We obtain the global fields (\ref{new_fields}) by numerically integrating
equations (\ref{eq1}) and using the stationary values of the dynamical
variables in (\ref{glob_field}) and these in (\ref{new_fields}). We then
construct bifurcation diagrams calculating, for each value of 
$\beta$, the fixed points of the system (\ref{new_var_eq}). Since each
node has an associated degree $k_i$, and, therefore, an associated
$\beta$, it is possible to project the bifurcation diagram in the
stationary pattern that resulted of the numerical integration of
(\ref{eq1}). The projection of the bifurcation diagram relative to the
variable $u_-$ on the stationary pattern for the difference
$u^{(1)}-u^{(2)}$ is shown in figure \ref{fig6}.

\end{appendix}

\end{document}